\documentclass[aps, 10pt,twocolumn,superscriptaddress,showpacs,preprintnumbers,amsmath,amssymb,prb,aps]{revtex4}

\usepackage{graphicx,bm}
\usepackage[dvips]{color}
\usepackage{multirow}

\newcommand{\SnInTe}{Sn$_{1-x}$In$_{x}$Te}
\newcommand{\angstrom}{\text{\normalfont\AA}}

\begin{document}

% Use the \preprint command to place your local institutional report
% number in the upper righthand corner of the title page in preprint mode.
% Multiple \preprint commands are allowed.
% Use the 'preprintnumbers' class option to override journal defaults
% to display numbers if necessary
%\preprint{}

%Title of paper
\title{Studies of the superconducting properties of Sn$_{\bm{1-x}}$In$_{\bm{x}}$Te ($\bm{x=0.38}$ to 0.45) using muon-spin spectroscopy}

\author{M. Saghir}
\email[]{M.Saghir@warwick.ac.uk}
\affiliation{Physics Department, University of Warwick, Coventry, CV4 7AL, United Kingdom}

\author{J. A. T. Barker}
\affiliation{Physics Department, University of Warwick, Coventry, CV4 7AL, United Kingdom}

\author{G. Balakrishnan}
\affiliation{Physics Department, University of Warwick, Coventry, CV4 7AL, United Kingdom}

\author{A. D. Hillier}
\affiliation{ISIS Facility, Science and Technology Facilities Council, Rutherford Appleton Laboratory, Chilton, Oxfordshire, OX11 0QX, U.K.}

\author{M. R. Lees}
\email[]{M.R.Lees@warwick.ac.uk}
\affiliation{Physics Department, University of Warwick, Coventry, CV4 7AL, United Kingdom}

\date{\today}

\begin{abstract}
The superconducting properties of Sn$_{1-x}$In$_{x}$Te ($x=0.38$ to 0.45) have been studied using magnetization and muon-spin rotation or relaxation ($\mu$SR) measurements. These measurements show that the superconducting critical temperature $T_c$  of Sn$_{1-x}$In$_{x}$Te increases with increasing $x$, reaching a maximum at around 4.8~K for $x=0.45$.  Zero-field $\mu$SR results indicate that time-reversal symmetry is preserved in this material. Transverse-field muon-spin rotation has been used to study the temperature dependence of the magnetic penetration depth $\lambda\left(T\right)$ in the mixed state. For all the compositions studied, $\lambda\left(T\right)$ can be well described using a single-gap $s$-wave BCS model. The magnetic penetration depth at zero temperature $\lambda\left(0\right)$ ranges from 500 to 580~nm. Both the superconducting gap $\Delta(0)$ at 0~K and the gap ratio $\Delta(0)/k_BT_c$ indicate that Sn$_{1-x}$In$_{x}$Te ($x=0.38$ to 0.45) should be considered as a superconductor with intermediate to strong-coupling.  
\end{abstract}

\pacs{74.25.-q,	%Properties of superconductors
			%74.25.Bt,	%Thermodynamic properties			74.25.Dw,	%Superconductivity phase diagrams			74.25.F-, %Transport properties
			74.25.Ha	%Magnetic properties including vortex structures and related phenomena
      74.62.Dh, %Effects of crystal defects, doping and substitution      74.70.Ad,	% Metals; alloys and binary compounds (including A15, MgB2, etc.)
			76.75.+i	%Muon spin rotation and relaxation
           }
\keywords{Superconductivity, Topological Insulators, Topological Crystalline Insulators, Muon spin rotation and relaxation, Effects of doping.}
\maketitle

\section{INTRODUCTION}
The discovery of the three-dimensional topological insulators (TIs) Bi$_2$Se$_3$ and Bi$_2$Te$_3$, with surfaces that have a gapless metallic state that is protected by time-reversal symmetry, has generated considerable excitement.\cite{Hasan, Qi} Further interest has resulted from the work that showed that Cu$_x$Bi$_2$Se$_3$ could be made superconducting, although the studies of superconductivity in this material have sometimes been hampered by the inhomogeneity of the samples.\cite{Hor,Kriener,Das} The half-Heusler compounds YPtBi, LaPtBi, and LuPtBi, all of which have noncentrosymmetric crystal structures and strong spin-orbit coupling, have also been investigated as candidates for 3D topological superconductivity.\cite{Butch,Goll,Tafti}
 
The existence of a different class of materials called topological crystalline insulators (TCI), in which the mirror symmetry rather than time-reversal symmetry produces the topologically protected metallic surface states, was proposed by Fu.\cite{Fu} Subsequent experimental work, including angle-resolved photo emission spectroscopy (ARPES), has shown that SnTe exhibits all the required characteristics of this TCI state.\cite{Hsieh,Tanaka} 

SnTe crystallizes in the rock salt structure and is classified a TCI because it satisfies the conditions required for the mirror symmetry. It is a narrow band semiconductor and in the as-grown state it usually forms with a number of Sn vacancies and is perhaps better described as Sn$_{1-\delta}$Te where $\delta$ is around 1\%. The Sn vacancies introduce holes. When the level of vacancies reaches $10^{20}$~cm$^{-3}$, Sn$_{1-\delta}$Te is a superconductor, but with a superconducting critical temperature $T_c$ remaining below 0.3~K for hole carrier densities $p$ up to $2\times10^{21}$~cm$^{-3}$. 

$T_c$ can be enhanced by replacing Sn with In.\cite{Erickson}  This doping with In introduces one further hole per In atom. For $x$ less than some critical In doping level $x_c\approx~2$\%, or viewed another way, for a hole doping level $p$ less than a critical level $p_c\approx 5\times10^{20}$~cm$^{-3}$, \SnInTe\ undergoes a ferroelectric structural phase transition adopting a rhombohedral structure at low temperature, while above this critical In doping level $x \geq x_c$ these materials remain cubic to zero kelvin.\cite{Erickson} For In substitution levels of around 6\% in Sn$_{0.988-x}$In$_{x}$Te, the $T_c$ of these materials is around 2~K. Point contact spectroscopy and ARPES measurements on Sn$_{1-x}$In$_{x}$Te with low levels of In substitution ($x = 0.045$) show the signature of a topological surface state~\cite{Sato} and suggest that this low-carrier-density superconductor exhibits surface Andreev bound states, odd-parity pairing, and topological superconductivity.\cite{Sasaki2} More recently, it has been shown~\cite{Zhong,Balakrishnan} that higher levels of In substitution ($x\sim0.4$ to 0.45) in Sn$_{1-x}$In$_{x}$Te gives a superconductor with a transition temperature as high as $\sim4.5$~K.

Studies of the bulk characteristics of Sn$_{1-x}$In$_{x}$Te, a material in which the superconductivity emerges from a parent TCI material, are essential in order to develop a more complete understanding of this important new class of materials. To this end, we have used muon-spin rotation and relaxation, $\mu$SR, to investigate the superconducting properties of \SnInTe\ for $0.38 \leq x \leq 0.45$. This level of In substitution gives close to the optimum $T_c$ for this system. $\mu$SR is an ideal probe to study the superconducting state as it provides microscopic information on the local field distribution within the bulk of the sample. It can be used to measure the temperature and field dependence of the London magnetic penetration depth, $\lambda$, in the vortex state of type-II superconductors.~\cite{Sonier, Brandt} The temperature and field dependence of $\lambda$ can in turn provide detailed information on the nature of the superconducting gap. This technique can also be used to detect small internal magnetic fields associated with the onset of an unconventional superconducting state.\cite{Aoki, Luke, Hillier}

\section{EXPERIMENTAL DETAILS}

\subsection{Sample preparation}
Samples of  \SnInTe\ for $x=0.38$, 0.40, 0.42, and 0.45, were prepared by the modified Bridgman method adopting a similar procedure to that described by Tanaka \textit{et al}.~\cite{Tanaka} for SnTe. Stoichiometric ratios of the starting materials, 99.99\% Sn,(shot) In (shot) and Te (powder), were placed in evacuated and sealed quartz ampoules. The quartz tubes were then heated to around $900~^{\circ}$C and slowly cooled ($2~^{\circ}$C/h) to $770~^{\circ}$C,  followed by a fast cooling to room temperature.   

Powder x-ray diffraction on powdered portions of the as grown boules were carried out using a Panalytical X' Pert Pro system with monochromatic Cu$K_{\alpha1}$ radiation.

Measurements of dc magnetization $M$ as a function of temperature $T$ at fixed applied field $H$, were made using a Quantum Design Magnetic Property Measurement System (MPMS) SQUID magnetometer, while measurements of the dc magnetization as a function of applied magnetic field at fixed temperature were carried out in an Oxford Instruments vibrating sample magnetometer (VSM).  
 
\subsection{Muon-spin rotation and relaxation experiments}

Muon-spin rotation and relaxation experiments were performed on the MuSR spectrometer of the ISIS pulsed muon facility, Rutherford Appleton Laboratory, UK. At ISIS, a pulse of muons with a full-width at half maximum (FWHM) of $\sim70$~ns is produced every 20~ms. The 100\% spin-polarized muons are implanted into a sample and after coming to rest the muon spin precesses in the local magnetic environment. The muons, with an average lifetime of $2.2~\mu$s, decay emitting a positron preferentially in the direction of the muon spin at the time of decay. 

In transverse-field (TF) mode, an external magnetic field was applied perpendicular to the initial direction of the muon spin polarization. The magnetic field was applied above the superconducting transition and the samples then cooled to base temperature (FC). In this configuration the signals from the instrument's 64 detectors were reduced to two orthogonal components which were then fitted simultaneously. Data were also collected in zero-field (ZF) mode. Here, the decay positrons from the muons were detected and time stamped in the detectors which are positioned either before $\left(B\right)$ or after $\left(F\right)$ the sample. The asymmetry in the positron emission as a function of time can then be determined as $A(t)=\left[N_B\left(t\right)-\alpha N_F\left(t\right)\right]/\left[N_B\left(t\right)+\alpha N_F\left(t\right)\right]$, where $N_F(t)$ and $N_B(t)$ are the counts in the forward and backward detectors respectively, and $\alpha$ is a relative counting efficiency for the forward and backward detectors.\cite{Lee:book,Yaouanc:book} In ZF mode, any stray fields at the sample position are canceled to within 10~$\mu$T by three pairs of coils forming an active compensation system.

The powdered samples of \SnInTe\ were mounted on high-purity silver plates. For each sample, the powder was packed into a circular indentation $\sim1$~mm deep and $\sim700$~mm$^2$ in area, cut into the square plate. In order to aid thermal contact, a small amount of General Electric (GE) varnish diluted with ethanol was added to the samples and the mixture allowed to dry.  In the TF mode, any silver exposed to the muon beam gives a non-decaying sinusoidal signal, while in the ZF relaxation experiments, any muons stopped in the silver sample holder give a time independent background. The $x=0.38$ and 0.45 samples were covered with thin silver foil and measured in an Oxford Instruments He$^3$ sorption cryostat with a base temperature of 0.30~K. The intermediate compositions with $x=0.40$ and 0.42 were covered with Al foil and were studied in a conventional Oxford Instruments He$^4$ cryostat with a base temperature of 1.1~K.

\section{EXPERIMENTAL RESULTS AND DISCUSSION}
\subsection{Structural Characterization}
The observed powder x-ray diffraction patterns collected at room temperature for the \SnInTe\ are consistent with the published patterns for the SnTe ($x=0$) parent phase and the In doped \SnInTe\ materials.\cite{Zhigarev,Zhong} The materials all have a cubic $Fm\bar{3}m$ structure. The lattice parameters calculated from the data are shown in Table~\ref{table_of_latticeparameters}. These values agree well with the published results for \SnInTe\ which show that there is a linear decrease in the lattice parameter $a$ with increasing $x$ until at around $x=0.40$ to 0.45, $a$ becomes almost constant with $x$.\cite{Zhigarev,Zhong} All our samples contain a small fraction of the tetragonal InTe phase, and the level of this phase increases with $x$. Once again this is consistent with previous published work which showed compositions around $0.3\leq x\leq0.5$ contain both the cubic and tetragonal materials, while for $x\geq0.6$ only the tetragonal phase of \SnInTe\ is formed.\cite{Zhong}

\begin{table}%-----------TABLE1--------------------------
\begin{center}
\begin{tabular}[t]{ll}\hline\hline
$x$ ~~& $a\left(\angstrom\right)$~~~\\\hline
0.38 ~~~& 6.2808(9) \\
0.40 ~~~& 6.280(1) \\
0.42 ~~~& 6.2798(6) \\
0.45 ~~~& 6.275(1) \\ \hline\hline
\end{tabular}
\end{center}
\caption{Room temperature lattice parameter $a$ determined from the powder x-ray diffractograms for the samples of \SnInTe\ with $x=0.38$, 0.40, 0.42, and 0.45. The samples are all cubic (space group $Fm\bar{3}m$).}
\label{table_of_latticeparameters}
\end{table}

\subsection{Superconducting properties}
\begin{figure}[tb]%-----------FIG1--------------------------
\begin{center}
\includegraphics[width=0.9\columnwidth]{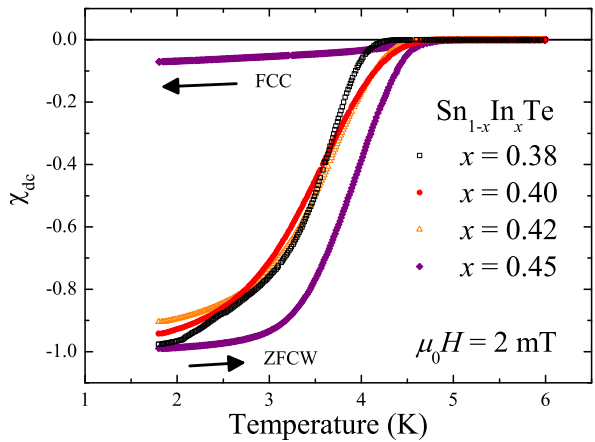}
\caption{\label{Figure1Saghir} (Color online). (a) Temperature dependence of the dc magnetic susceptibility for four samples of \SnInTe\ with $x=0.38$, 0.40, 0.42, and 0.45 measured on zero-field-cooled warming (ZFCW) in an applied field of 2~mT. The field-cooled cooling (FCC) curve for the $x=0.45$ is also shown. The FCC data for the other three samples are similar and are omitted for clarity. A demagnetization factor has been applied to account for the shapes of the samples.~\cite{Aharoni}}
\end{center}
\end{figure}

Measurements of the dc magnetic susceptibility ($\chi_{\rm{dc}}=M/H$) as a function of temperature were used to investigate the onset of superconductivity. $\chi_{\rm{dc}}(T)$ measurements were carried out in applied fields of 2~mT. Figure~\ref{Figure1Saghir} shows the zero-field-cooled warming (ZFCW) data for all four samples we have studied and the field-cooled cooling (FCC) data for the sample with $x=0.45$. The transition temperature of each sample is taken as the temperature at which we observe the onset of a clear (5\% of the full Meissner signal) diamagnetic signal. $T_c$ increases slightly with increasing $x$ saturating at around 4.80(5)~K for $x=0.45$, in good agreement with of previous studies of the doping dependence of $T_c$.~\cite{Zhong,Balakrishnan} The transitions are all rather broad and the magnetization observed in the superconducting state only begins to saturate at the lowest temperatures measured. We also note that a very small diamagnetic signal extends up to higher temperature (e.g. up to $T=5.14(2)$~K in the case of the $x=0.45$ sample). Figure~\ref{Figure1Saghir} also shows the field-cooled cooling data for one of the four samples. On cooling, there is an expulsion of the flux at the same temperature as on heating. The low temperature diamagnetic signal is around one tenth of that seen in the ZFCW data at 1.8~K indicating considerable bulk pinning of the magnetic flux.

The lower critical fields $H_{c1}$ estimated from the first deviation from linearity in the low-field regions of our magnetization versus applied field $M\left(H\right)$ scans (data not shown) are of the order of 1 to 2~mT at 1.8~K.

\begin{figure}[tb]%-----------FIG2--------------------------
\begin{center}
\includegraphics[width=0.9\columnwidth]{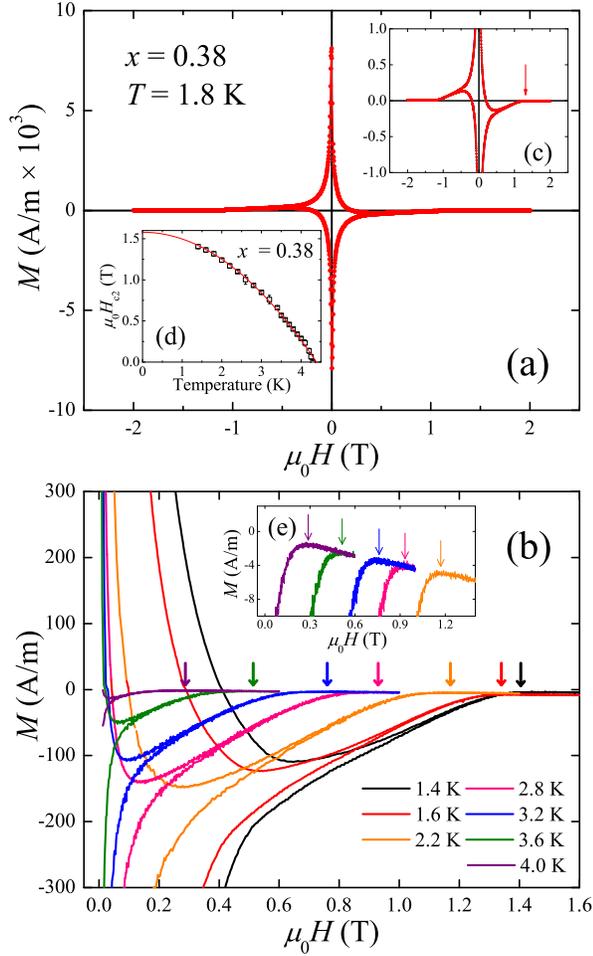}
\caption{\label{Figure2Saghir} (Color online). (a) Magnetization versus applied field $M\left(H\right)$ loop collected at 1.8~K for Sn$_{0.62}$In$_{0.38}$Te. (b) Two quadrant $M\left(H\right)$ loops for the same Sn$_{0.62}$In$_{0.38}$Te sample collected at different temperatures. A demagnetization factor has been applied to account for the shape of the sample.~\cite{Aharoni} The inset of Fig.~\ref{Figure2Saghir}(c), shows the $M\left(H\right)$ curve collected at 1.8~K plotted on an expanded $y$ scale with the upper critical field $H_{c2}$ (also see Fig.~\ref{Figure2Saghir}(d)) marked with an arrow.  The inset Figure~\ref{Figure2Saghir}(e) shows several of the other $M\left(H\right)$ loops enlarged around zero magnetization. These data were used to estimate the temperature dependence of the upper critical field $H_{c2}\left(T\right)$ as shown in the inset, Fig.~\ref{Figure2Saghir}(d).}
\end{center}
\end{figure}

High field $M\left(H\right)$ scans can be used to extract values for $H_{c2}(T)$. A typical four-quadrant $M\left(H\right)$ loop at 1.8~K for a sample with $x=0.38$ is shown in Fig.~\ref{Figure2Saghir}(a). Two-quadrant $M\left(H\right)$ loops at selected temperatures up to $T_c$ for the same sample are shown in Fig.~\ref{Figure2Saghir}(b). The $M\left(H\right)$ loops all exhibit hysteresis at low fields indicating significant bulk pinning of the flux-line lattice. As the field increases the loops close. At 1.8~K the loop is reversible above 1~T, however, $H_{c2}$ in not indicated by this closing of the loop but by a sharp change in slope at 1.31(5)~T. The  $\mu_0H_{c2}(T)$ values  for Sn$_{0.62}$In$_{0.38}$Te extracted from these data are shown in Fig.~\ref{Figure2Saghir}(d). $H_{c2}(T)$ is fitted using a simple parabolic temperature dependence, $H_{c2}(T) = H_{c2}(0)\left[1-\left(T/T_c\right)^2\right]$, to estimate the zero temperature limit of the upper critical field $\mu_0H_{c2}(0)$. The values of $\mu_0H_{c2}(0)$ for the $x=0.38$ as well as the other compositions are given in Table~\ref{table_of_gapparameters} and range from 1.56(1)~T for $x=0.38$ to 1.62(1)~T for $x=0.45$. The $H_{c2}(T)$ curves for the four samples are later used to correct the values of $\lambda\left(T\right)$ determined from the muon spectroscopy data.

\subsection{Transverse-field muon-spin rotation}

\begin{figure}[tb!]%-----------FIG3--------------------------
\begin{center}
\includegraphics[width=0.9\columnwidth]{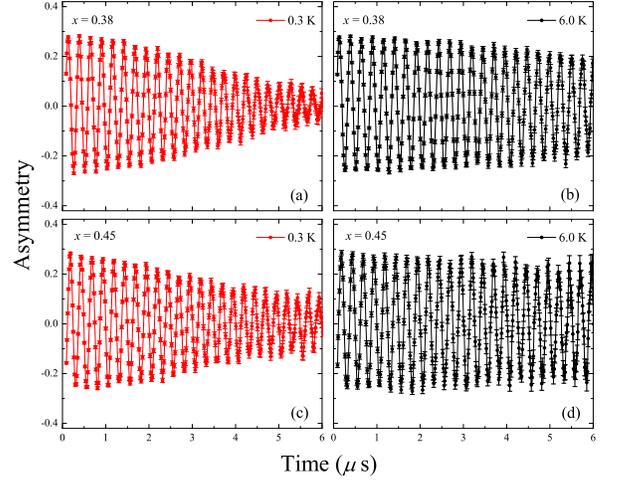}
\caption{\label{Figure3Saghir} (Color online) Transverse-field asymmetry spectra (one component) collected in an applied magnetic field of 30~mT at (a) 0.30 and (b) 6.0~K for the $x=0.38$, and (c) 0.30 and (d) 6.0~K for the $x=0.45$ phases of Sn$_{1-x}$In$_{x}$Te.}
\end{center}
\end{figure}

TF-$\mu$SR asymmetry data were collected at different temperatures for each of the four samples studied. The samples were cooled in fixed transverse-fields of between 5 and 50~mT. Our $M\left(H\right)$ data indicate that in these fields, below $T_c$, the samples are in the mixed state. The temperature was then increased in steps through $T_c$. At each temperature step a TF-$\mu$SR time spectrum containing at least $20\times10^6$ muon decay events was collected. Figure~\ref{Figure3Saghir} shows the TF-$\mu$SR precession signals in 30~mT, collected below and then above $T_c$ for the $x=0.38$ and $x=0.45$ compositions, and are typical of the data collected during this study. 

\begin{figure}[tb!]%-----------FIG4--------------------------
\begin{center}
\includegraphics[width=0.9\columnwidth]{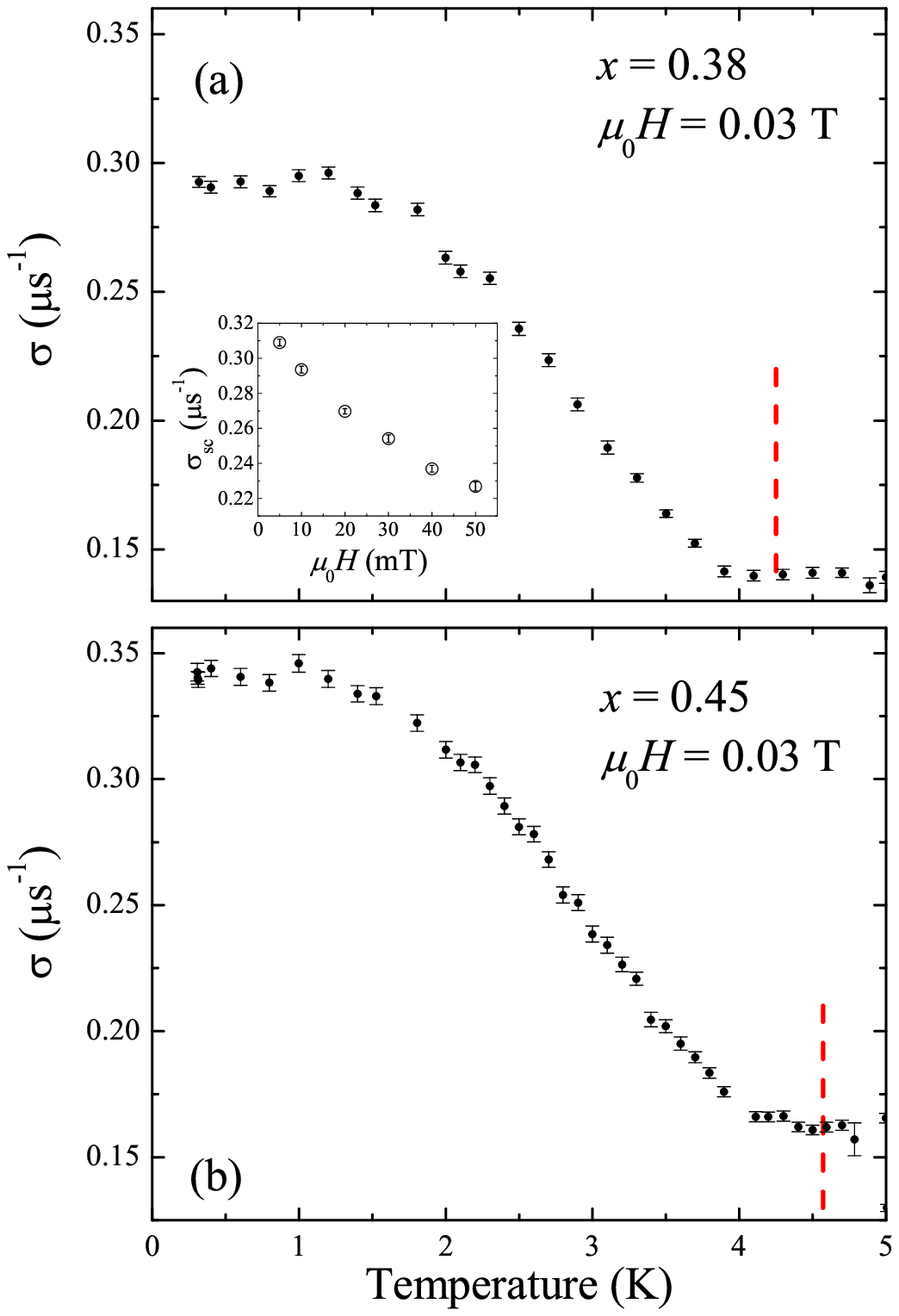}
\caption{\label{Figure4Saghir}(Color online) Temperature dependence of the muon-spin relaxation rate $\sigma$, collected in an applied magnetic field $\mu_0H=0.03$~T for (a) $x=0.38$ and (b) $x=0.45$ compositions of Sn$_{1-x}$In$_{x}$Te. The vertical dashed lines indicate the $T_c$ as determined from magnetic susceptibility data. The inset in the upper panel shows the magnetic field dependence of $\sigma_{\textnormal{sc}}$, obtained at 0.30~K for the sample with $x=0.38$.}
\end{center}
\end{figure}

Figures~\ref{Figure3Saghir}(b) and ~\ref{Figure3Saghir}(d) show that in the normal state ($T>T_c$), the signal decays slowly, due to the homogeneous magnetic field distribution throughout the samples. In contrast, Figs.~\ref{Figure3Saghir}(a) and ~\ref{Figure3Saghir}(c) show that in the superconducting state ($T<T_c$), the rate at which the asymmetry of the signal is lost is much more rapid, due to the inhomogeneous field distribution arising from the flux-line lattice. The TF-$\mu$SR precession data were fitted using an oscillatory decaying Gaussian function,

\begin{eqnarray}
\label{Depolarization_Fit}
G_X(t)=A_{1}\exp\left(-\sigma^{2}t^{2}\right/2)\cos\left(\omega_{1} t +\phi\right)  \nonumber \\
+A_{2}\cos\left(\omega_{2} t +\phi\right),~
\end{eqnarray}

\noindent where $\omega_1$ and $\omega_2$ are the frequencies of the muon precession signal and the background signal respectively, $\phi$ is the initial phase offset, and $\sigma$ is a Gaussian muon-spin relaxation rate.  Figures~\ref{Figure4Saghir}(a) and ~\ref{Figure4Saghir}(b) show the temperature dependence of $\sigma$ obtained from the TF-$\mu$SR data collected in an applied of 30~mT for the $x=0.38$ and 0.45 phases of Sn$_{1-x}$In$_{x}$Te. 

The relaxation rate $\sigma$ can be written as $\sigma=\left(\sigma^{2}_{\textnormal{sc}} + \sigma^{2}_{\textnormal{nm}}\right)^{\frac{1}{2}}$, where $\sigma_{\textnormal{sc}}$ is the superconducting contribution to the relaxation rate and $\sigma_{\textnormal{nm}}$ is the nuclear magnetic dipolar contribution which is assumed to be constant over the temperature range of the study. The inset in Fig.~\ref{Figure4Saghir}(a) shows the magnetic field dependence of $\sigma_{\textnormal{sc}}$ at 0.30~K for the $x=0.38$ sample. The data confirm that the magnetic field dependence of $\sigma_{\textnormal{sc}}$ must be considered when these data are used to extract values of the magnetic penetration depth $\lambda$.  

At some temperature $T<T_c$ in a superconductor with an upper critical field $B_{c2}(T)$ that is not many times larger than the internal field $B$ and in which there is a hexagonal Abrikosov vortex lattice, the muon-spin depolarization rate $\sigma_{\textnormal{sc}}(T)$ is related to the penetration depth $\lambda(T)$ at the same temperature by~\cite{Brandt2003}

\begin{equation}
\label{BrandtEq}
\begin{split}
\sigma_{\mathrm{sc}}\left(T\right)\left[\mu\textnormal{s}^{-1}\right]=4.83 \times10^4\left(1-\frac{B}{B_{c2}\left(T\right)}\right) \\ 
\times\left[1+1.21 \left(1-\sqrt{\frac{B}{B_{c2}\left(T\right)}}\right)^3\right]\lambda^{-2}\left(T\right)\left[\textnormal{nm}\right].
\end{split}
\end{equation}

Therefore, using the values of $B_{c2}\left(T\right)$ from magnetization and $\sigma_{\textnormal{sc}}\left(T\right)$ from $\mu$SR we have determined the temperature dependence of the London magnetic penetration depth $\lambda\left(T\right)$. The penetration depth $\lambda\left(T\right)$ can then be fit within the local London approximation~\cite{Tinkham,Prozorov} for an $s$-wave BCS superconductor in the clean limit using the expression

\begin{equation}
\left[\frac{\lambda^{2}\left(0\right)}{\lambda^{2}\left(T\right)}\right]_{\rm{clean}}=1+2\int^{\infty}_{\Delta\left(T\right)}\left(\frac{\partial f}{\partial E}\right)\frac{ EdE}{\sqrt{E^2-\Delta^2\left(T\right)}},
\end{equation}

\noindent where $f=\left[1+\exp\left(E/k_BT\right)\right]^{-1}$ is the Fermi function and $\Delta\left(T\right)=\Delta_{0}\delta\left(T/T_c\right)$. The temperature dependence of the gap is approximated by the expression~\cite{Tinkham} $\delta\left(T/T_c\right)=\tanh\left\{1.82\left[1.018\left(T_c/T-1\right)\right]^{0.51}\right\}$.

Alternatively in the dirty limit we have 
\begin{equation}
\left[\frac{\lambda^{2}\left(0\right)}{\lambda^{2}\left(T\right)}\right]_{\rm{dirty}}=\frac{\Delta\left(T\right)}{\Delta\left(0\right)}\tanh\left[\frac{\Delta\left(T\right)}{2k_BT}\right].
\end{equation}

\begin{figure}[tb!]%-----------FIG5--------------------------
\begin{center}
\includegraphics[width=0.9\columnwidth]{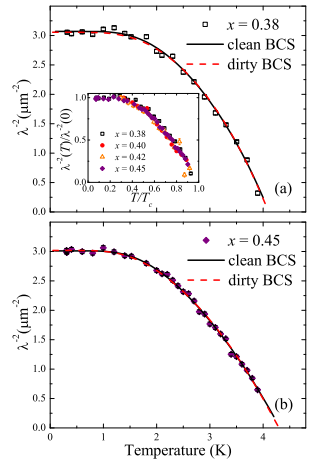}
\caption{\label{Figure5Saghir} (Color online) Inverse square of the London penetration depth $\lambda^{-2}$ as a function of temperature for (a) $x=0.38$ and (b) $x=0.45$ compositions of Sn$_{1-x}$In$_{x}$Te. Fits to the data in the BCS clean and dirty limit are indicated by the solid and dashed lines respectively. The inset shows ${\lambda^{-2}\left(T\right)}/{\lambda^{-2}\left(0\right)}$ as a function of the reduced temperature, $T/T_c$, for all four compositions of Sn$_{1-x}$In$_{x}$Te studied.}
\end{center}
\end{figure}

We obtain good fits to the $\lambda^{-2}\left(T\right)$ data for all four samples using both the models discussed above (see Fig.~\ref{Figure5Saghir}). There is little difference between the quality of the fits, as measured by $\chi^2_{\textnormal{norm}}$, in the clean and dirty limits. The superconducting parameters, including the magnitude of the gap, $T_c$, and the penetration depth $\lambda\left(0\right)$ at $T=0$~K determined from these fits are given in Table~\ref{table_of_gapparameters}.  As expected, the magnitudes of the gap in the clean limit are larger than those obtained for the dirty limit (e.g., for $x=0.38$ $\Delta$ is reduced to 0.66(3)~meV in the dirty limit), but in both cases the values obtained place the materials in the intermediate to strong-coupling limit. There is reasonable quantitative agreement between the penetration depths calculated from our $\mu$SR measurements and the values of $\lambda\left(0\right)$ determined previously from dc magnetic susceptibility data.\cite{Balakrishnan} The transition temperatures determined from the bulk muon spectroscopy data are systematically lower than those given by the magnetization measurements. This may, in part, be due to the lower applied fields used in the magnetization measurements, and the requirement to establish an inhomogeneous field distribution in order to depolarize the muons. Any small differences between the transition temperatures determined from fits to the $\lambda^{-2}\left(T\right)$ data and those estimated from the $\sigma\left(T\right)$ data are due to uncertainties in establishing exactly where $\sigma=\sigma_{\textnormal{nm}}$ and the determination of $\lambda\left(T\right)$ from $\sigma_{\textnormal{sc}}\left(T\right)$ data made using Eq.~\ref{BrandtEq}.

\begin{table}[ht]%-----------TABLE2--------------------------
\begin{center}
\begin{tabular}[t]{lllll}\hline\hline
~~& $x=0.38$ & $x=0.40$ & $x=0.42$~~~&$x=0.45$ \\\hline
$T_c^{\textnormal{mag}}$ (K) ~~~& 4.25(5)& 4.69(5)& 4.57(8) & 4.80(5) \\
$B_{c2}(0)$ (T) & 1.56(1)& 1.59(1)&1.58(1) & 1.62(1) \\
$\lambda(0)$~(nm)~~& 572(7)&542(5) & 496(4) & 578(2) \\
$T_c^{\mu\textnormal{SR}}$~(K)~~& 4.08(8)& 4.5(1)&4.20(8) & 4.30(2) \\
$\Delta(0)$~(meV)~~& 0.73(3)&0.67(2) & 0.67(3) & 0.70(1) \\
$\Delta(0)/k_{B}T_c^{\mu\textnormal{SR}}$ ~& 2.08(9)&1.73(6) &1.85(9) & 1.89(3) \\ \hline\hline
\end{tabular}
\caption{Superconducting parameters extracted from the magnetization data and from the fits to the TF muon spectroscopy data for all four compositions of \SnInTe\ studied. $T_c^{\textnormal{mag}}$ is the superconducting critical temperature determined from the magnetization versus temperature curves, while $T_c^{\mu\textnormal{SR}}$, the penetration depth $\lambda\left(0\right)$, and the gap $\Delta\left(0\right)$ are extracted from fits to the $\lambda^{-2}\left(T\right)$ data in the clean limit. The upper critical field values $B_{c2}\left(0\right)$ were determined from $M\left(H\right)$ loops collected at fixed temperature.}
\label{table_of_gapparameters}
\end{center}
\end{table} 

Plotting ${\lambda^{-2}\left(T\right)}/{\lambda^{-2}\left(0\right)}$ as a function of the reduced temperature, $T/T_c$, for the different compositions of Sn$_{1-x}$In$_{x}$Te reveals that all the data collected fall on a single universal curve (see inset of Fig.~\ref{Figure5Saghir}). This scaling of the data clearly suggests that in this composition range at least, \SnInTe\ has the same gap symmetry.

We now compare these results with previous work on \SnInTe. To date, there have only been a limited number of studies of the bulk superconducting properties of \SnInTe\ for $0.3 \leq x \leq 0.5$. Low-temperature measurements of the thermal conductivity, $\kappa$, for a single crystal of Sn$_{0.6}$In$_{0.4}$Te with a $T_c$ of 4.1~K gave a small value for the residual thermal conductivity $\kappa_0/T$ in zero magnetic field and weak field dependence for this term.\cite{He} It was argued that this gave strong evidence for a full (nodeless) superconducting gap in the bulk although it was also suggested that the superconducting state was most likely an unconventional odd-parity state. The results are also consistent with estimates of the gap ratio $\alpha=\Delta\left(0\right)/k_BT_c$, a measure of the coupling strength, for Sn$_{0.6}$In$_{0.4}$Te made using the temperature dependence of the heat capacity $C\left(T\right)$ below $T_c$, although an estimate for the strength of the coupling from the jump in $C\left(T\right)$ at $T_c$ gives lower values.\cite{Balakrishnan} 

The results obtained in our study are in line with previous work on more lightly doped \SnInTe.\cite{Erickson, Novak} Heat capacity data for \SnInTe\ with 0.02 $\leq x \leq 0.08$ were fitted using a modified BCS theory which allows $\alpha$ to vary.\cite{Novak,Padamsee} For $x=0.025$, the hole doping $p$ is less than $p_c$, and \SnInTe\ is fully gapped with $\alpha=1.77$ and a $T_c$ of 1.44~K. For slightly higher doping levels ($x=0.04$, $p=5\times10^{20}$cm$^{-3}$, $T_c=1.2$~K and 0.05 with $p=7\times10^{20}$cm$^{-3}$, $T_c=1.41$~K) the fits gave $\alpha$ values of 1.83 and 1.88 respectively which are are larger, but the $C\left(T\right)$ data are still well described by a modified BCS model.\cite{Novak} In an earlier study of materials in this doping regime, estimates of the coupling strength made from heat capacity data collected below $T_c$ also suggested these materials were superconductors with intermediate to strong coupling. However, in the same way as for the more heavily doped Sn$_{0.6}$In$_{0.4}$Te, the $\alpha$ values estimated from the jump in $C\left(T\right)$ at $T_c$ give weaker coupling values.\cite{Erickson}

\begin{figure}[tb!]%-----------FIG6--------------------------
\begin{center}
\includegraphics[width=0.9\columnwidth]{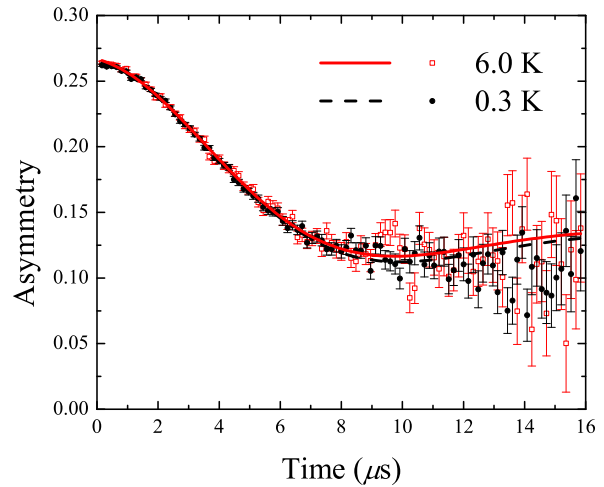}
\caption{\label{Figure6Saghir} (Color online) ZF-$\mu$SR time spectra collected at 6~K (open symbols) and 0.30~K (closed symbols) for the $x=0.45$ sample of Sn$_{1-x}$In$_{x}$Te. Fits to the data using the Kubo-Toyabe function as described in the text are shown as solid (6~K) and dashed (0.3~K) lines.}
\end{center}
\end{figure}

\subsection{Zero-field muon-spin relaxation}

We have also performed ZF-$\mu$SR measurements on the $x=0.45$ composition of Sn$_{1-x}$In$_{x}$Te in order to search for any (weak) internal magnetism that may arise as a result of ordered magnetic moments, as well as to look for any temperature dependent relaxation processes associated with the onset of superconductivity.~\cite{Aoki, Luke, Hillier} In these relaxation experiments, any muons stopped in the silver sample holder give a time independent background. The ZF-$\mu$SR time spectra at base temperature (0.30~K) and 6~K are shown in Fig.~\ref{Figure6Saghir}. There is no precessional signal, ruling out the possibility of a large internal field and hence long-range magnetic order. These data can be modeled using

\begin{equation}
\label{KT2}
G\left(t\right)=A_{0}G^{\textnormal{KT}}\left(t\right)\exp\left(-\Lambda t\right)+A_{\textnormal{bkgd}},
\end{equation} 
\noindent where the depolarization can be described by the Kubo-Toyabe function,~\cite{Hayano}
\begin{equation}
\label{KT1}
G^{\rm{KT}}\left(t\right)=\left[\frac{1}{3}+\frac{2}{3}\left(1-\sigma_\textnormal{{KT}}^{2}t^{2}\right)\exp\left(-\frac{\sigma_\textnormal{{KT}}^{2}t^{2}}{2}\right)\right],
\end{equation}  
\noindent where $A_{0}$ is the initial asymmetry, $A_{\rm{bkgd}}$ is the background, $\sigma_\textnormal{{KT}}$ is the relaxation rate associated with the nuclear moments and $\Lambda$ is the electronic relaxation rate. The fits yield the parameters shown in Table~\ref{ZF_table_of_fits} that to within error are the same above and below $T_{c}$, confirming the qualitative view that there are no changes in the form of the data with temperature. Note that fits with $\Lambda$ fixed at zero also gave values for $A_{0}$, $\sigma_\textnormal{{KT}}$, and $A_{\rm{bkgd}}$ above and below $T_c$ that agree to within error.

\begin{center}%-----------TABLE3--------------------------
\begin{table}[ht]
\begin{tabular}[t]{lllll}\hline\hline
$T\left(\textnormal{K}\right)$ & $A_{0}$&$\sigma_\textnormal{{KT}}$($\mu$s$^{-1})$ & $\Lambda$($\mu$s$^{-1})$& $A_\textnormal{{bkgd}}$ \\\hline
6.0 & 0.154(2) & 0.179(4) & 0.038(6) & 0.113(2)\\
0.3 & 0.157(2) & 0.176(3) & 0.034(4) & 0.108(2)\\\hline\hline
\end{tabular}
%\par\medskip
\caption{Parameters extracted from the fits using the Kubo-Toyabe function to the zero-field $\mu$SR data collected above and below $T_{c}$ for \SnInTe\ with $x=0.45$ shown in Fig.~\ref{Figure6Saghir}.}
\label{ZF_table_of_fits}
\end{table}
\end{center}

The behavior observed in the ZF-$\mu$SR data, and the values of $\sigma_\textnormal{{KT}}$ extracted from the fits, are commensurate with the presence of random local fields arising from the nuclear moments within the samples that are static on the time scale of the muon precession. There is no evidence for any additional relaxation channels that may be associated with more exotic electronic phenomena such as the breaking of time-reversal symmetry.~\cite{Aoki, Luke, Hillier} 

\section{SUMMARY}
We have performed $\mu$SR and magnetization studies on bulk polycrystalline samples of \SnInTe\ for $0.38 \leq x \leq 0.45$.  The materials are all bulk superconductors with $T_c$ increasing with $x$ up to a maximum of around 4.80~K for $x=0.45$. The upper critical fields are estimated to be $\sim1.6$~T in this doping range. Magnetization versus field loops indicate bulk pinning in low applied fields while the response is close to reversible in applied fields close to $H_{c2}$. There is no evidence in these materials either for long-range magnetic order or for any unusual electronic behavior in the superconducting state. Our $\mu$SR  results also confirm that time-reversal symmetry is preserved in this system. The absolute values of the magnetic penetration depth $\lambda\left(0\right)$ lie in the range 500 to 580~nm. The temperature dependence of $\lambda$ for \SnInTe\ in this doping range can be adequately described using a single isotropic $s$-wave gap; bulk \SnInTe\ appears to be fully gapped superconductor. The magnitude of the superconducting gap ratio $\alpha$ for all four samples suggest that these are intermediate to strong-coupling superconductors. 

\begin{acknowledgments}
This work was supported by the EPSRC, UK (EP/I007210/1). Some of the equipment used in this research was obtained through the Science City Advanced Materials project: Creating and Characterizing Next Generation Advanced Materials project, with support from Advantage West Midlands (AWM) and part funded by the European Regional Development Fund (ERDF). We wish to thank D. Walker and S. York for assistance with the x-ray and compositional analysis of the samples used in this study.   
\end{acknowledgments}

% Create the reference section using BibTeX:
\bibliography{Saghir_v1}

\end{document}